\documentstyle{article}
\begin{document}
\hoffset -1.5cm
\large

 %Inserted by TeXtelmExtel

\title{\Large\bf On Charge Quantization and Abelian Gauge Horizontal 
Symmetries}
\author{\large L.N. Epele$^1$, C.A. Garc\'\i a Canal$^1$, and William A.
Ponce$^2$\\
\normalsize 1-Laboratorio de F\'\i sica Te\'orica, Departamento de F\'\i sica \\
\normalsize Universidad Nacional de La Plata, C.C. 67-1900 La Plata-Argentina \\
\normalsize 2-Departamento de F\'\i sica, Universidad de Antioquia\\
\normalsize A.A. 1226, Medell\'\i n, Colombia}

 %Inserted by TeXtelmExtel

\maketitle

\begin{center}
Acepted for publication on Phys. Letters {\bf B}.
\end{center}

 %Inserted by TeXtelmExtel

\begin{abstract}
Under the assumption that there exists a local gauge horizontal symmetry 
$G_H$ wich allows only for a top quark mass at tree level, we look for the 
constraints that charge quatization and the family structure of the 
standard model imposes on that symmetry.
\end{abstract}

 %Inserted by TeXtelmExtel

\renewcommand{\baselinestretch}{1.2}
\setlength{\hsize}{16cm}
\setlength{\vsize}{12in}

 %Inserted by TeXtelmExtel

\section{Introduction}
Among the many questions that the successful Standard Model (SM) leaves
unanswered, one of the most basic is why the gauge group is 
$G_{SM}=SU(3)_c\otimes SU(2)_L\otimes U(1)_Y$. In fact, given $N_F$ 
Weyl fermion fields, the kinetic energy term, with the convention 
of taking all fields to be left-handed, 
possesses a global $U(N_F)$ symmetry; then why does Nature gauges only 
$G_{SM}$ which is a tiny subgroup of this enormous global group? Certainly there
is a reason why 
the entire $U(N_F)=SU(N_F)\otimes U(1)$ group cannot be gauged: 
the resulting theory would be anomalous\cite{zee}.
Nevertheless, groups larger than $G_{SM}$ should in principle be allowed.

 %Inserted by TeXtelmExtel

Looking the problem upside down, one notices that the SM $U(1)_Y$ hypercharge assignments, 
for the standard fermions, are least to say, peculiar\cite{wilc}.
This observation leads us to ask a simpler question:
given $SU(3)_c\otimes SU(2)_L$ as a local gauge group with the 
standard matter spectrum (supported by experimental facts) 
\begin{equation}\label{spe}
 Q_L\sim (3,2); \hspace{.5cm} U^c_L\sim (\bar{3},1); 
\hspace{.5cm} D^c_L\sim (\bar{3},1); 
\hspace{.5cm} F_L\sim (1,2); 
\hspace{.5cm} E_L\sim (1,1) 
\end{equation}
(where $(k,l)$ refers to $(SU(3)_c,SU(2)_L)$ labels), which is the 
most general $U(1)$ group associated with hypercharge that can 
be gauged together with $SU(3)_c\otimes SU(2)_L$? 
Is the answer the standard $U(1)_Y$ with the mentioned "peculiar''
assignment of hypercharge quantum numbers?

 %Inserted by TeXtelmExtel

The purpose of this note is precisely to specify the minimal set of conditions that selects
the standard model hypercharges. Our analysis is mainly based on anomaly
cancellation considerations and includes horizontal symmetries that 
when broken, give rise to a mass hierarchy where only the top quark acquire
mass at the tree level.
In going forward with our analysis we will recall well known results relevant 
for our purposes.

 %Inserted by TeXtelmExtel

Notice that $SU(3)_c\otimes SU(2)_L$ gauge theory with the spectrum
referred in Eq.(\ref{spe})
is free of the Adler-Bell-Jackiw anomaly\cite{abj} because
$SU(2)$ is real (pseudoreal), and the standard matter
spectrum is vector-like with respect to $SU(3)_c$.
Also it is free of the Global $SU(2)_L$ anomaly\cite{witen} because the
number of $SU(2)_L$ doublets is even. Consequently, anomalies
could come only from extra $U(1)$ gauge groups.

 %Inserted by TeXtelmExtel

The first step is to look for all the $U(1)$'s that could be included 
under general conditions.
For $3$ families, the most general symmetry of the fermion kinetic 
energy which commutes with $SU(3)_c\otimes SU(2)_L$ is: 
\[ G=U(3)_Q\otimes U(3)_U\otimes U(3)_D\otimes U(3)_F\otimes U(3)_E
\equiv[U(3)]^5 \]
where $U(3)_\eta = SU(3)_\eta\otimes U(1)_\eta ( \eta= Q,U,D,F,E)$ and
$U(1)_\eta$ is a global (family independent) abelian factor. Again, the
entire group $G$ can not be gauged because it is anomalous, but which are 
the abelian subgroups of $G$ that can be gauged so that the resulting theory
will be anomaly free? If they are family independent, in the sense that the 
same hypercharge value is assigned to related multiplets of each 
 one of the three families, they must be subgroups of

 %Inserted by TeXtelmExtel

\[U(1)_Q\otimes U(1)_U\otimes U(1)_D\otimes U(1)_F\otimes U(1)_E
\equiv [U(1)]^5. \]

 %Inserted by TeXtelmExtel

 If they assign different hypercharge values to related multiplets in each 
family, they must be subgroups of

 %Inserted by TeXtelmExtel

\[SU(3)_Q\otimes SU(3)_U\otimes SU(3)_D\otimes SU(3)_F\otimes SU(3)_E
\equiv [SU(3)]^5. \]

 %Inserted by TeXtelmExtel

Let us present the analysis of a family independent $U(1)$ abelian factor. 
In this case the anomaly cancellation constraint equations are\cite{several}:

 %Inserted by TeXtelmExtel

\begin{eqnarray}
\label{a}  {[SU(2)]^2U(1) }& : &  3Y_Q+Y_F=0 \\* 
\label{b}{[SU(3)]^2U(1) }& : &  2Y_Q+Y_U+Y_D=0 \\*
\label{c}{[grav]^2U(1)  }& : &  6Y_Q+3Y_U+3Y_D+2Y_F+Y_E=0 \\* 
\label{d}{[U(1)]^3      }& : &  6Y_Q^3+3Y_U^3+3Y_D^3+2Y_F^3+Y_E^3=0 
\end{eqnarray}

 %Inserted by TeXtelmExtel

\noindent
where $Y_\eta ( \eta=Q,U,D,F,D)$ are the $U(1)$ hypercharges for the corresponding 
multiplet. Here we have included the constraint comming from the mixed
gauge-gravitational anomaly\cite{ds} resulting from triangle diagrams
involving two energy momentum tensors (gravitons) vertices and a $U(1)$
gauge vertex.

 %Inserted by TeXtelmExtel

Equations (\ref{a})-(\ref{d}) have the following three different solutions:

 %Inserted by TeXtelmExtel

\noindent

 %Inserted by TeXtelmExtel

\[ {\bf Sol. A} \hspace{1cm}  
Y_F=-3x, \, Y_E=6x, \, Y_Q=x, \, Y_U=-4x, \, Y_D=2x \]

 %Inserted by TeXtelmExtel

\[{\bf Sol. B} \hspace{1cm}
Y_F=-3y, \, Y_E=6y, \, Y_Q=y, \, Y_U=2y, \, Y_D=-4y \]

 %Inserted by TeXtelmExtel

\[{\bf Sol. C} \hspace{3.5cm}
Y_F=Y_E=Y_Q=0, \; Y_U=-Y_D=z \]

 %Inserted by TeXtelmExtel

Where $x,y$ and $z$ are arbitrary real parameters, fixed by
normalization. Solution {\bf A} with $x=1/3$ yields the usual $U(1)_Y$
hypercharge of the SM ($x=-1/3$ is related to mirror families), while
solutions {\bf B} and {\bf C} represent two different $U(1)$ 
quantum number assignments, not identified with conventional physics. 

 %Inserted by TeXtelmExtel

Solution {\bf B} appears as a consequence of the symmetry 
$Y_U\leftrightarrow Y_D$ in the anomaly constraint equations, which means 
that from the anomaly cancellation alone one can not distinguishes between  
$U$ and $D$ quark fields. So, solution {\bf A} 
and solution {\bf B} became equivalent when $Y_U$ and $Y_D$ are interchanged, 
but this would be not the case if an extra-interaction, which distinguishes between 
$U$ and $D$, is introduced\cite{pecei}. Since we wish to analize 
horizontal symmetries which could perform this distinction, we treat the three solutions 
independently. Then, anomaly cancellation by itself is not enough
to fix completely the standard gauge group\cite{several}. 
In fact, this ambiguity is absent in the SM standard formulation because the
Higgs sector of the 
theory plays a central role. The introduction of a Higgs field $\phi_{SM}$ with $U(1)$
hypercharge $Y_\phi$ which couples to Up and Down quarks and to charged 
leptons in each family, reduces 
the number of free hypercharge parameters in the analysis via the 
relationships: 

 %Inserted by TeXtelmExtel

\begin{equation}
Y_Q+Y_U=-Y_Q-Y_D=-Y_F-Y_E=Y_\phi.
\end{equation}

 %Inserted by TeXtelmExtel

\noindent
and the solution {\bf A} is unambigously singled out\cite{several}.
Notice that the SM Higgs provides mass to every fermion field. This means that
the phenomenological quark and lepton mass spectrum hierarchy should appear from
a fine tunning of the Yukawa couplings, turning the scheme quite unnatural. 

 %Inserted by TeXtelmExtel

There is also an alternative view of that problem, used in the 
last two papers of Ref.\cite{several}. There one defines an electric charge 
operator $Q_{EM}=T_{3L}+Y/2$, with the extra assumption that the $U(1)_{EM}$ 
associated with $Q_{EM}$ is vectorlike. 

 %Inserted by TeXtelmExtel

 Before exploring the possibilities to single out solution {\bf A}, 
we can pursue
the analysis in an almost general case, taking into account all three solutions, 
namely $U(1)_{Y_A}, U(1)_{Y_B}$ and $U(1)_{Y_C}$ respectively. 
In this context, one can ask oneself whether 
$U(1)_{Y_A}\otimes U(1)_{Y_B}\otimes U(1)_{Y_C}$ can be gauged simultaneusly, or at least
two of them at the same time. The answer is {\bf no} because 
the triangle anomalies
$[U(1)_{Y_\alpha}]^2U(1)_{Y_\beta}$ with $\alpha, \beta = A,B$ or $C$ and
$\alpha \ne\beta$,
do not cancel. So, the hypercharges $Y_B$ and $Y_C$
should be automatically excluded
if $U(1)_{Y_A}$ is gauged as the correct hypercharge.
However, the cancellation of the triangle anomalies allows that
once one gauges solution {\bf A} for a
particular value of $x$, one can gauge as many $U(1)_{Y_A}$ as one wishes,
each one for a different value of $x$. In other words, the only
family independent hypercharge
that one can gauge simultaneously with $G_{SM}$ is a $U(1)_{Y^\prime}$ of 
hypercharge $Y^\prime\sim Y_A$ \cite{ponce,pzw}.

 %Inserted by TeXtelmExtel

In summary, from anomaly cancellation considerations alone, one can specify
as a sensible gauge goup: $SU(3)_c \otimes SU(2)_L \otimes U(1)$, but with an ambiguity in 
the assignement of the abelian charge, namely hypercharges $Y_A,\,Y_B$ or $Y_C$.
In what follow we analyse the uselfulness of the inclusion of a further symmetry related to families, namely a local gauge horizontal symmetry $G_H$ in selecting the SM "peculiar" hypercharges.
In so doing we ask, as a further requirement, that the spontaneous breaking of the symmetry 
provides mass, at the tree level, only to the top quark. This last condition sounds clearly
reasonable whenever one recalls that $m_b / m_t \sim 3\%$, i.e., of the same order as the amount of quantum (radiative) corrections.   

 %Inserted by TeXtelmExtel

\section{Charge Quantization from Horizontal Symmetries}
 In this section we introduce an Abelian gauge horizontal symmetry for three 
families designed to allow only for a tree level top quark mass. The generation of 
masses for the remaining known fermion fields via radiative corrections demands 
for a ``Diophantine solution"\cite{dioph} to the new anomaly contraint 
equations. We then show that the implementation of this program is compatible 
only with solution {\bf A} of the previous section, ruling out solutions 
{\bf B} and {\bf C}, implying thus discrete values for the electric charge.

 %Inserted by TeXtelmExtel

\subsection{Horizontal symmetries}
Since there is no evidence for an specific local gauge 
horizontal symmetry, we work within the following frame 
which is at least  consistent with experimental facts:
There exist only three complete chiral families 
of ordinary matter (without right-handed neutrinos) togheter with a local 
gauge group $SU(3)_c\otimes SU(2)_L\otimes U(1)_Y\otimes G_H$, where 
$G_H\subset [U(3)]^5$, the most general symmetry allowed. 
 This scheme is implemented with the following symmetry breaking chain 

 %Inserted by TeXtelmExtel

\[SU(3)_c\otimes SU(2)_L\otimes U(1)_Y\otimes G_H
{\longrightarrow}SU(3)_c\otimes SU(2)_L\otimes U(1)_Y
{\longrightarrow}SU(3)_c\otimes U(1)_{EM}\] 

 %Inserted by TeXtelmExtel

\noindent
We also expect that an effective Lagrangean of the type\cite{ellis} 
\begin{eqnarray}
{\cal L}^{eff}&=\sum_{i,j}[(y^U_{ij}\phi^\dagger_{SM}U^c_{iL}
\left({\Lambda \over M}\right)^{n^U_{ij}}
\left({\Lambda^\star \over M}\right)^{n^{\prime U}_{ij}}
+ y^D_{ij}\stackrel{\sim}{\phi}^\dagger_{SM}D^c_{iL}
\left({\Lambda \over M}\right)^{n^D_{ij}}
\left({\Lambda^\star \over M}\right)^{n^{\prime D}_{ij}})Q_{jL}  \nonumber \\ 
& + y^E_{ij}\stackrel{\sim}{\phi}^\dagger_{SM}E^c_{iL}F_{jL} 
\left({\Lambda \over M}\right)^{n^E_{ij}}
\left({\Lambda^\star \over M}\right)^{n^{\prime E}_{ij}}] + h.c. 
\end{eqnarray}
will provide, after the symmetry breaking and diagonalization, the appropriate 
masses and mixing angles for all the ordinary fields. In ${\cal L}^{eff}, \,\Lambda$ 
is the breaking scale of $G_H$, $M$ is an undetermined mass scale, $n^\kappa_{ij}$ and $n^{\prime\kappa}_{ij}, 
(\kappa=U,D,E)$ are integer numbers 
for $i,j=1,2,3$ and $y^\kappa_{ij}$ are Yukawa coupling constants of order one. 
To properly fulfil all the requirements in ${\cal L}^{eff}$,one needs
a $G_H$ able to distinguish Up from Down, and integer $G_H$ quantum numbers to 
relate them to the loop exponentials $n^\kappa_{ij}$ and 
$n^{\prime\kappa}_{ij}$. 

 %Inserted by TeXtelmExtel

The simplest local gauge horizontal symmetries contained in $[U(3)]^5$
that one may consider are $U(1)_H$ and $U(1)_{H_1}\otimes U(1)_{H_2}$. 
That they are reasonable horizontal symmetries for understanding
the mass spectrum of the elementary fermions and their mixings has been
disscused in Refs.\cite{pzw,ibanez}. In what follows we are going to restrict 
ourselves to those two cases\cite{zepe}. Notice that these symmetries could be realized
both in a family independent and in a family dependent way. As we have stated above, 
this means that the same horizontal charge is assigned to related 
multiplets of each family or not.

 %Inserted by TeXtelmExtel

\vskip4mm
\subsection{Family Independent}
\vskip 3mm
\noindent
We consider first a family independent $U(1)_H$ as our horizontal
symmetry group.
For a Higgs field with $U(1)_H$ charge $Y_\phi$, a Yukawa coupling for the
top quark is allowed if $Y_{Q_3}+Y_{U_3}=Y_\phi$, whereas a bottom quark
coupling is forbidden if $Y_{Q_3}+Y_{D_3}\neq -Y_{\phi}$ which is in
contradition with equation anomaly constraint equivalent to Eq. ({\ref{b}}). Therefore for a $U(1)_H$ family independent, 
if a top quark mass arises at tree level, a bottom mass arises as well at
the same level\cite{pzw}. So, only a $U(1)_H$ family dependent symmetry could be
consistent with our working hypothesis. A similar analysis and conclusions
follow for a family independent $U(1)_{H_1}\otimes U(1)_{H_2}$. 

 %Inserted by TeXtelmExtel

\vskip4mm
\subsection{Family Dependent}
\vskip 3mm
\noindent
For three families, an $U(1)_H$ family dependent symmetry group 
must be a subgroup of
$[SU(3)]^5$. Consequently, for each horizontal $\eta$ multiplet ($\eta = Q,U,D,F,E$), the
hypercharges must be traceless, and even further they must be of the form
$(\delta_\eta ,0, -\delta_\eta )$ or
$(\delta_\eta ,\delta_\eta , -2\delta_\eta)$. These are the forms of the 
corresponding $U(1)$ subgroups of $SU(3)$. Then, for the most general
$U(1)_H$ extracted from $[SU(3)]^5$, we must have
\[trace [U(1)_H]=\sum_{i=1}^3 H_{\eta_i}=0\]
for $\eta =Q,U,D,F,E$ and where $H_{\eta_i}$ stand for the horizontal charges.

 %Inserted by TeXtelmExtel

Now, if the anomalies are going to be cancelled by an interplay among
the three families, the linear cancellation constraints
$[SU(2)_L]^2U(1)_{H}$, $[SU(3)_c]^2U(1)_{H}$, $[grav]^2U(1)_{H}$, and 
$[U(1)]^2U(1)_{H}$ are automatically satisfied, and we have to worried
only about the following two new constraints:
                                                             
 %Inserted by TeXtelmExtel
                                                             
\begin{eqnarray}
\label{h21} {[U(1)_H]^2U(1) } & : &  \sum_i (6Y_QH^2_{Q_i}+3Y_UH^2_{U_i}+
                             3Y_DH^2_{D_i}+2Y_FH^2_{F_i}+Y_EH^2_{E_i})=0 \\*
{[U(1)_{H}]^3   } & : &  \sum_i (6H_{Q_i}^3+3H_{U_i}^3+3H_{D_i}^3
                                     +2H_{F_i}^3+H_{E_i}^3)=0
\end{eqnarray}
where $i=1,2,3$ sums over the multiplets for the three families, and the
$Y_\eta$ hypercharge for $\eta=Q,U,D,F,E$, appearing in Eq.(\ref{h21}), are in principle any 
one of the three possible solutions {\bf A}, {\bf B}, or {\bf C}. 

 %Inserted by TeXtelmExtel

For an $U(1)_H$ of the form $(\delta_\eta,0,-\delta_\eta )$, the quadratic
equation (\ref{h21}) becomes: 

 %Inserted by TeXtelmExtel

\begin{equation}
6Y_Q\delta^2_Q + 3Y_U\delta^2_U + 3Y_D\delta^2_D + 
2Y_F\delta^2_F + Y_E\delta^2_E=0  \label{quad}, 
\end{equation}
and the cubic equation is trivially satisfied. 

 %Inserted by TeXtelmExtel

For an $U(1)_H$ of the form $(\delta_\eta,\delta_\eta,-2\delta_\eta)$, the 
quadratic equation is again given by Eq.(\ref{quad}) and the cubic 
equation reads: 
\begin{equation}\label{cub}
6\delta_Q^3 + 3\delta_U^3 + 3\delta_D^3 + 2\delta_F^3 + \delta_E^3 = 0, 
\end{equation}
\noindent
which do not depend upon the $U(1)$ hypercharge $Y_\eta$, so it does not 
play any role on the solution of the ambiguity in the $U(1)$ hypercharge 
(eventhough it may play an important role on the form of the mass matrices). 

 %Inserted by TeXtelmExtel

Now, since at tree level the mass matrix for the Up quark sector must be a 
rank one matrix, we may, without lost of generality, relabel the weak 
eigenstates in such a way that only the ($3,3$) entry in the Up quark mass 
matrix is different from zero. Then it is just natural to demand for a 
symmetric mass matrix for the entire Up quark sector\cite{ibanez}, which is 
achieved only if $\delta_Q=\delta_U(\equiv \delta)$. 

 %Inserted by TeXtelmExtel

On the other hand, 
 since Eq. (\ref{quad}) is also $U\leftrightarrow D$ symmetric, in order to 
distinguishes Up from Down let us 
look for a $G_H=U(1)_{H_U}\otimes U(1)_{H_D}$ where $U(1)_{H_U}$ is 
inert (zero charge value) in the Down sector and $U(1)_{H_D}$ is inert 
in the Up sector. Then for a $G_H=U(1)_{H_U}\otimes U(1)_{H_D}$ and a 
symmetric Up quark mass matrix we have the following anomaly constraint 
quadratic equations:
\begin{eqnarray}
{[U(1)_{H_U}]^2U(1)} &:& 3\delta^2(2Y_Q+Y_U)+2Y_F\delta_F^2 
+ Y_E\delta_E^2=0 \label{aa} \\
{[U(1)_{H_D}]^2U(1)} &:& 3Y_D\delta^{\prime 2}_D+2Y_F\delta_F^{\prime 2} 
+ Y_E\delta_E^{\prime 2}=0 \label{bb} \\
{[U(1)_{H_U}]^2U(1)_{H_D}} &:& 2\delta^2_F\delta_F^\prime 
+ \delta^2_E\delta_E^\prime=0\\ \label{cc}
{[U(1)_{H_D}]^2 U(1)_{H_U}} &:& 2\delta_F\delta_F^{\prime 2} 
+ \delta_E\delta_E^{\prime 2}=0 \label{dd}
\end{eqnarray}
\noindent
where the unprimed charges are related to the $U(1)_{H_U}$ factor and the 
primed charge to the $U(1)_{H_D}$ factor. 
 A Diophantine analysis\cite{dioph} of the last two 
equations shows that it is impossible to find rational value for 
$\delta_F , \delta^\prime_F, \delta_E , \delta^\prime_E$ which satisfies 
them. Since irrational charges are not allowed in our approach, we must 
ask for non-overlapping $U(1)_{H_U}$ and $U(1)_{H_D}$ factors, which in turn 
implies that the last two equations must be trivially satisfied (0=0).

 %Inserted by TeXtelmExtel

If $Y_\eta$ are given by solution {\bf C}, Eqs. (\ref{aa}) 
and (\ref{bb}) 
are satisfied only for $\delta=\delta^\prime_D=0$. This solution will 
produce tree level masses in the Up and Down sectors (which can be seen from 
the matrix quantum numbers for the Up and Down quark sectors presented 
below) and must be ruled out. 

 %Inserted by TeXtelmExtel

On the other hand, if $Y_\eta$ is given by solution {\bf B}, Eqs. (\ref{aa}) and (\ref{bb}) become:
\begin{eqnarray}
2\delta^2-\delta^2_F+\delta^2_E&=&0 \\ 
2\delta^{\prime 2}_D+\delta^{\prime 2}_F-\delta^{\prime 2}_E&=&0 
\end{eqnarray}
Non-overlapping real solutions are produced only for 
$\delta_E=\delta^\prime_F=0$. But for those values there is no rational 
solution. Then, solution {\bf B} is also ruled out. 

 %Inserted by TeXtelmExtel

If $Y_\eta$ is given by solution {\bf A}, equations (\ref{aa}) and (\ref{bb}) 
become: 
\begin{eqnarray}
\delta^2+\delta^2_F-\delta^2_E&=&0 \\ 
\delta^{\prime 2}_D-\delta^{\prime 2}_F+\delta^{\prime 2}_E&=&0,
\end{eqnarray}
\noindent
where non-overlaping real solutions exist for 
$\delta_F=\delta^\prime_E=0$. Those solutions are $\delta_E=\pm\delta$ and 
$\delta_F^\prime=\pm\delta_D^\prime$. 

 %Inserted by TeXtelmExtel

Consequently,  only an $U(1)_Y$ with the hypercharge given by solution {\bf A} is 
consistent with our working hypothesis, toghether with 
a $U(1)_{H_U}\otimes U(1)_{H_D}$ local gauge horizontal symmetry which 
assigns charge values to ordinary fields according to the values in 
Table I. Since the two factors $U(1)_{H_U}$ and $U(1)_{H_D}$ do not overlap, 
$U(1)_{H_U}\otimes U(1)_{H_D}$ is equivalent to a single 
$U(1)_H$ with hypercharge values as in Table I.

 %Inserted by TeXtelmExtel

\vskip3mm

 %Inserted by TeXtelmExtel

 %Inserted by TeXtelmExtel

 %Inserted by TeXtelmExtel

\section{Mass matrices} 
Let us briefly comment on the mass matrices produced by the simplest solution 
of the anomaly constraint equations, presented in Table {\bf I}.

 %Inserted by TeXtelmExtel

\subsection{Case $U(1)_H$ of the form $(\delta_\eta, 0, -\delta_\eta)$}
\vskip 2mm

 %Inserted by TeXtelmExtel

For an $U(1)_H$ of this form, Eq.(\ref{quad})with the hypercharge values 
$Y_\eta$ given by solution {\bf A} reads:
\begin{equation}
\label{cquad}
6x(\delta_Q^2-2\delta_u^2+\delta_D^2-\delta_F^2+\delta_E^2)=0
\end{equation}
which is satisfied by the $U(1)_H$ charge values of Table I as it 
should be (although a more general Diophantine analysis of this equation may 
be performed). Since the 
cubic equation is trivially satisfied by a $U(1)_H$ of the form 
$(\delta_\eta, 0, -\delta_\eta)$, there are no further constraints for the 
$U(1)_H$ charge values in Table I. For those values the $U(1)_H$ matrix 
quantum numbers, for the Up quark sector, is given by

 %Inserted by TeXtelmExtel

\[\left(\begin{array}{ccc}
2\delta   &  \delta  &   0    \\
\delta    &    0     & -\delta\\
0         &  -\delta &   -2\delta
\end{array} \right). \]

 %Inserted by TeXtelmExtel

\noindent
Similarly, for the Down quark sector  we obtain 

 %Inserted by TeXtelmExtel

\[\left(\begin{array}{ccc}
\delta+\delta^\prime   &  \delta  &   \delta-\delta^\prime    \\
\delta^\prime    &    0     & -\delta^\prime     \\
-\delta+\delta^\prime    &  -\delta &  -\delta-\delta^\prime
\end{array} \right). \]

 %Inserted by TeXtelmExtel

Then, a $\phi_{SM}$ Higgs field with a $U(1)_H$ charge value $-2\delta$ 
will produce, at tree level, a rank one mass matrix for the Up quark sector 
$(M^{up}_{33}\sim m_t$ as wished) 
and a zero mass matrix for the Down quark sector, as far as 
$\delta^\prime\neq\pm\delta,\pm 2\delta,\pm 3\delta$. 

 %Inserted by TeXtelmExtel

Now, under the (crude) assumption that $y^\eta_{ij}=1$ for 
$\eta=U,D,E\, ;ij=1,2,3$, the effective Lagrangean produces, 
for $\delta^\prime=-5\delta$ and the appropriate $U(1)_H$ charge value 
for the necessary Higgs, the following quark mass matrices\cite{ibanez}:

 %Inserted by TeXtelmExtel

\[M_U\sim\left(\begin{array}{ccc}
\theta^2   &  0         &    \theta    \\
0          & \theta     &       0      \\
\theta     &  0         &       1 
\end{array} \right), \]
and 
\[M_D\sim\left(\begin{array}{ccc}
\theta^3     &  0         & \theta^2     \\
0            & \theta     & 0            \\
\theta^4     &  0         & \theta
\end{array} \right). \]
Where $\theta=\Lambda/M$. The diagonalization of $M_U$ and $M_D$ 
produces $m_t\sim 1+\theta^2, m_c\sim \theta, m_u\sim 0, 
m_b\sim\theta + \theta^5, m_s\sim\theta$ and $m_d\sim\theta^3-\theta^5$. These 
values  may be brought closer to the experimental 
results by the appropriate selection of $\theta$ and $y_{ij}(\sim1)$ values. 

 %Inserted by TeXtelmExtel

Notice by the way that the matrix quantum numbers for the charged lepton 
sector is equivalent to the matrix quantum numbers for 
the Down quark sector, which in turn implies a relationship between charged 
leptons and Down quark masses as it is usually expected. 

 %Inserted by TeXtelmExtel

\vskip 3mm
\subsection{Case $U(1)_H$ of the form $(\delta_\eta,\delta_\eta,-2\delta_\eta)$}
\vskip 2mm

 %Inserted by TeXtelmExtel

For $U(1)_H$ of this form, both equations (\ref{cub}) and (\ref{cquad}) 
must be be satisfied. As mentioned above, Eq.(\ref{cquad}) is satisfied by 
the $U(1)_H$ charge values in Table I, and a Diophantine solution to 
Eq.(\ref{cub}) brings the further constraints 

 %Inserted by TeXtelmExtel

\[\delta_E=-\delta; \delta_D=-\delta_F=-2\delta \]

 %Inserted by TeXtelmExtel

\noindent
then the $U(1)_H$ matrix quantum numbers for the Up quark sector for 
this case reads
\[\left(\begin{array}{ccc}
2\delta   &  2\delta  & -\delta    \\
2\delta   &  2\delta & -\delta\\
-\delta   &  -\delta &   -4\delta
\end{array} \right), \]

 %Inserted by TeXtelmExtel

\noindent
and for the down quark sector it reads
\[\left(\begin{array}{ccc}
-\delta    & -\delta   & 5\delta  \\
-\delta    & -\delta   & 5\delta\\
-4\delta   &  -4\delta & 2\delta
\end{array} \right); \]

 %Inserted by TeXtelmExtel

\noindent
then a $\phi_{SM}$ Higgs field with a $U(1)_H$ charge value $-4\delta$ 
will produce also a rank one mass matrix at tree level for the Up quark sector 
(again $M^{up}_{33}\sim m_t)$ and a zero mass matrix for the Down quark sector. 
The analysis of the mass matrices for the simplest solutions presented in this 
case does not improve the results obtained for the previous case. 

 %Inserted by TeXtelmExtel

\section{Conclusions}
 The previous analysis allows us to conclude that:\\
- In order to provide a tree level mass only for the top quark, at least an Abelian
horizontal family dependent symmetry $U(1)_H$ is needed.\\
- The Abelian horizontal family dependent symmetry can be only of the form 
$(\delta_\eta, 0 -\delta_\eta)$ or $(\delta_\eta,\delta_\eta,-2\delta_\eta)$ or 
a combination of both (for $\eta=Q,U,D,E,F$).\\
- The presence of a horizontal gauge symmetry, allowing only for a top quark
mass at the tree level, selects the Standard Model hypercharges among the three different
weak hypercharge assignments compatible with anomaly cancellation. As a matter 
of fact, the
very simple election of $G_H = U(1)_{H_U}\otimes U(1)_{H_D}$ (equivalent
to $U(1)_H$), with the horizontal charges summarized in Table I, does
the job, breaking simultaneously the $Y_U\leftrightarrow Y_D$ symmetry.

 %Inserted by TeXtelmExtel

In our approach the tree level ``top" mass, generated by the Yukawa coupling 
with an appropriate Higgs boson, is expected to be the seed mass that gives rise
to the three fermion mass matrices via quantum effects.

 %Inserted by TeXtelmExtel

In summary, a realistic Standard Model in the sense of observing the
experimental mass hierarchy, could emerge from the inclusion of a horizontal
local gauge symmetry. Moreover, the proposed mechanism based on anomaly
cancellation, is sufficently restrictive as to uniquely fix the standard
weak hypercharges.

 %Inserted by TeXtelmExtel

A further analysis of the mass matrices for more general Diophantine solutions 
of the anomaly constraint Eq.(\ref{quad}), and for a general $U(1)_H$ of the 
form $(\delta_\eta,0-\delta_\eta)\oplus (\delta_{\eta^\prime}, 
\delta_{\eta^\prime}, -2\delta_{\eta^\prime}),\, \eta\neq\eta^\prime$, is 
under way.

 %Inserted by TeXtelmExtel

\noindent
{\bf Acknowledgements:}
This work was partially supported by CONICET, Argentina and COLCIENCIAS,
Colombia. W.A.P acknowledges the hospitality of the Physics Department
of the Universidad de La Plata in Argentina during the completition of 
this work.

 %Inserted by TeXtelmExtel

\vspace{3cm}

 %Inserted by TeXtelmExtel

{\bf Table I}. Hypercharge values for $U(1)_{H_U}, U(1)_{H_D}$ and $U(1)_H$ 
local gauge horizontal symmetries. 

 %Inserted by TeXtelmExtel

\begin{tabular}{l|ccccc} \hline
            & $\delta_Q$ & $\delta_U$ & $\delta_D$ & $\delta_F$ 
            & $\delta_E$ \\ \hline\hline
$U(1)_{H_U}$ & $\delta$ & $\delta$ & 0 & 0 & $\pm\delta$ \\
$U(1)_{H_D}$ & 0 & 0 & $\delta^\prime$ & $\pm\delta^\prime$ & 0 \\
$U(1)_H$     & $\delta$ & $\delta$ & $\delta^\prime$ & $\pm\delta^\prime$ & 
               $\pm\delta$ \\ \hline
\end{tabular}

 %Inserted by TeXtelmExtel

\pagebreak

 %Inserted by TeXtelmExtel

 %Inserted by TeXtelmExtel


\begin{thebibliography}{99}
\bibitem{dioph}
The Diophantine analysis is a branch of the theory of numbers 
related to the solution of a system of algebraic equations using only 
entire numbers and their powers.
\bibitem{zepe}
In the second paper of Ref.\cite{pzw} it was shown that these are the only two 
possible local gauge groups consistent with the working hypothesis 
stated.
\end{thebibliography}
\end{document}